\documentclass[aps,prl,groupedaddress]{revtex4}

\usepackage{graphicx}

\begin{document}
 
\title{On the generation and the nonlinear dynamics of X-waves of the Schroedinger equation}
\author{Claudio Conti}
\email{c.conti@ele.uniroma3.it}
\homepage{http://optow.ele.uniroma3.it}
\affiliation{
NOOEL, Nonlinear Optics and OptoElectronics Laboratory,\\
National Institute for the Physics of Matter, INFM - University Roma Tre,
Via della Vasca Navale 84, 00146 Rome - Italy}
%
\date{\today}
\begin{abstract}
The generation of finite energy packets of X-waves is analysed in 
normally dispersive cubic media by using an X-wave expansion.
The 3D nonlinear Schroedinger model is reduced
to a 1D equation with anomalous dispersion.
Pulse splitting and beam replenishment as
observed in experiments with water and Kerr media 
are explained in terms of a higher order breathing soliton. 
The results presented also hold in periodic media and Bose-condensed gases.  
\end{abstract}
%
\maketitle
%
%
\section{Introduction}

X-waves, originally introduced by Lu and Greenleaf in acoustics,\cite{Lu92,Lu92a} can be simply described as a non-monochromatic superposition of 
plane-waves with the same wavevector component along
a given direction of propagation.
The resulting pulsed beam can travel undistorted without neither diffraction or dispersion, 
even in the absence of a nonlinear self-action.
Such X packets have been the subject of intense research in several fields, 
as recently reviewed in \cite{Salo00,Saari01,Recami03}.

The first experimental evidence of X-wave formation during nonlinear
optical processes was reported in \cite{QELS01}. 
Subsequently, X-related results have appeared in 
\cite{DiTrapani03,Jedrkiewicz03,DiTrapani04,Faccio2004}, including
a theoretical analysis and discussion on the generation and existence 
of this class of spatially and temporally localized waves in 
nonlinear media \cite{Conti03}.

The above mentioned experimental and theoretical results have stimulated further investigations on the role of 
X-waves in nonlinear processes, as well as their description
in terms of linear waves with a convenient envelope approach,
\cite{Trillo02,Orlov02,Orlov02a,Butkus03, Conti03b,Porras03,Porras03a,Porras03b,Ciattoni04,Longhi04,Longhi04a,Longhi04b}
and their extension to quantized fields. \cite{Conti04b}
Within the active field of optical self-invariant pulsed beams
(see among others \cite{Ziolkowski93,Saari97,Shaarawi97,Reivelt00,Sheppard02,Grunwald03,Ciattoni04b}),
it was recognized that X-waves may play a fundamental
role in all the nonlinear processes encompassing spatio-temporal mechanics,
well beyond the realm of quadratic frequency mixing where they 
had been originally observed. 
 X-waves have been also predicted in periodic media and Bose condensed gases. \cite{Conti04}

X- or \textit{progressive undistorted waves} in focusing or defocusing Kerr media can be associated to the instability 
of wide beams propagating in the presence of 
a refractive index linearly dependent on intensity.
\cite{Conti03c} 
This effect has also been investigated in quadratically nonlinear media
\cite{Conti04b,Longhi04}. 

Very recently light filamentation with femtoseconds pulses in water, \cite{Dubietis03}
was explained by the ``nonlinear X-wave paradigm''. \cite{Kolesik04}
The authors pointed out that these spatio-temporal packets
can be highly dynamic, i.e. they are continuously generated, undergoing splitting and replenishment with an ``average'' invariant propagation.
Such a picture points out the robustness of X-waves with respect to the
numerous non-trivial effects that may take place in 3+1D nonlinear processes. 
On the contrary, self-trapped light bullets, which do not exist
in the absence of a nonlinear response, are extremely sensitive 
to the specific model. 
X-wave dynamics seems strongly connected to 
pulse splitting during self-focusing in normally dispersive media, as originally predicted 
in 1992 \cite{Chernev92,Rothenberg92}, 
and later investigated by several authors. 
\cite{Ranka96,Berge96,Brabec97,Trippenbach97,Diddams98,Manassah98,
Litvak00,Berge02}
Experimental investigation of splitting/replenishment 
has been reported in \cite{DiTrapani03b}.

The generation of X-waves may also be taken as the basis for the interpretation
of the conical emission observed during the propagation of powerful ultrashort laser pulses in air, or water. 
\cite{Nibbering96,Kosareva97,Dubietis03,Zeng04}

In this manuscript a preliminary settlement of X-wave nonlinear dynamics is attempted. 
While the forthcoming theory is developed with reference to nonlinear optics, as outlined in \cite{Conti04} it applies to Bose
Einstein condensation (i.e. to ``matter X-waves''), as well. 
The central idea is that, if these wave-packets can 
be considered somehow as ``modes of free space'', with a given direction
of propagation $z$, then their nonlinear evolution is essentially two dimensional,
involving $z$ and time.
Thus, under certain conditions, the problem can be reformulated and strongly
simplified by adopting the well known approach of guided wave nonlinear optics.

Once an appropriate X-wave expansion is determined, it is not straightforward
to write down ``coupled X-wave equations'', in analogy with coupled
mode equations, for the simple reason that these spatio-temporal beams are
not normalizable. The first step is then to build finite energy 
solutions, and then use the resulting superposition of X-waves to investigate nonlinear dynamics. 
Section I and II cover the early stages of this approach.
In section III nonlinear regimes are addressed via a perturbative expansion.
This is legitimated by the fact that progressive undistorted waves do exist
in the absence of nonlinearity.
The analysis clarifies in which sense X-waves are robust, and may constitute a ``paradigm''
for nonlinear 3D+1 dynamics.
In section IV I consider the highly nonlinear regime. This is possible while
limiting to a specific X-wave. If the wave packet
is spontaneously generated during a nonlinear process (e.g. from 
a bell shaped pulsed beam) the resulting picture, 
for example the emergence of breathing solutions,
provides a relevant insight for interpreting reported
numerical and experimental findings.
The last sections deal with some corollaries: 
compression and chirp of X-waves, and the relation with the integrable 1D nonlinear Schroedinger equation.
The appearance of an integrable model, enforcing the use of categories such as
``breathers'',
and its effectiveness in interpreting experimental and numerical investigations, 
points to an intriguing connection between nonlinear X-waves and solitons.
\section{X-waves for the 3D+1 Schroedinger equation}
The wave equation (at angular carrier frequency $\omega_0$) describing paraxial propagation in normally dispersive media,
at the lowest order of approximation, can be cast as
\begin{equation}
\label{NLSdimensional}
i\partial_z A+ik'\partial_T A+\frac{1}{2k}\nabla^2_{xy}A-\frac{k''}{2}\partial_{TT}A=0\text{,}
\end{equation}
where $k=\omega_0 n(\omega_0)/c$ and its derivatives provide the dispersion. 
In a reference system traveling at the group velocity of the medium, $t=T-k'z$, it is
\begin{equation}
\label{NLSdimensional2}
i\partial_z A+\frac{1}{2k}\nabla^2_{xy}A-\frac{k''}{2}\partial_{tt}A=0\text{.}
\end{equation}
For the sake of simplicity hereafter I will consider radially symmetric beams, with $r=\sqrt{X^2+Y^2}$,
.

\begin{figure}
\includegraphics[width=80mm]{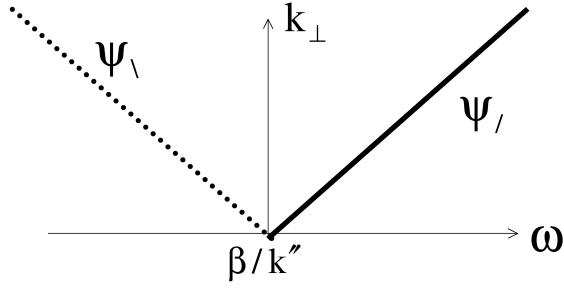}
\caption{\label{figurepsiX} Sketch of spatio-temporal spectra of ``slash'' and ``backslash'' X-waves.}
\end{figure}

Progressive undistorted waves, propagating with inverse differential velocity $\beta$ 
\cite{noteA1}
can be retrieved by looking for solutions
of the form $A=\psi(t-\beta z,r)\exp(ik_z z)$, thus ($\tau\equiv t-\beta z$)
\begin{equation}
\label{X1}
-k_z \psi-i\beta \partial_\tau \psi-\frac{k''}{2}\partial_\tau^2\psi+\frac{\nabla^2_{xy}}{2 k}\psi=0\text{.}
\end{equation}
If $\psi$ is written as a superposition of monochromatic Bessel beams, $J_0(\sqrt{k k''}\alpha r)\exp(-i\omega\,\tau)$, with $\alpha$ in frequency units, the corresponding spatio-temporal dispersion relation is
\begin{equation}
-k_z-\beta\omega+\frac{k''}{2}\omega^2=\frac{k'' \alpha^2}{2}\text{.}
\end{equation}
In order to have a continuous spectrum along $\omega$, the left-hand side must be positive, thus ensuring the absence
of evanescent waves. 
This can be achieved, in the simplest way, by letting $k_z=-\beta^2/2 k''$, which gives
\begin{equation}
\label{dispersionX}
(\omega-\frac{\beta}{k''})^2=\alpha^2\text{.}
\end{equation}
Eq. (\ref{dispersionX}) implies the existence of two types of X-waves:
\begin{equation}
\psi_\slash(\tau ,r,\beta)=\int_{0}^\infty e^{-i(\frac{\beta}{k''}+\alpha)\tau}J_0(\sqrt{k''k}\alpha r)f_\slash(\alpha)d\alpha
\end{equation}
and 
\begin{equation}
\psi_\backslash(\tau ,r,\beta)=\int_{0}^\infty e^{-i(\frac{\beta}{k''}-\alpha)\tau}J_0(\sqrt{k''k}\alpha r)f_\backslash(\alpha)d\alpha\text{,}
\end{equation}
with the corresponding ``spectra'' $f_\slash(\alpha)$ and $f_\backslash(\alpha)$.
They will be denoted as ``slash'' and ``backslash'' X-waves, because of 
the shape of their spatio-temporal frequency content, as
 discussed below (see Fig.\ref{figurepsiX}). 

A general linear X-wave solution,\cite{noteX} traveling with inverse differential 
velocity $\beta$, is given by
\begin{equation} 
\begin{array}{l}
A_X=e^{-i\frac{\beta^2}{2k''}z}[\psi_\slash(t-\beta z,r,\beta)+\psi_\backslash(t-\beta z,r,\beta)]\text{,}
\end{array}
\end{equation}
which can be rewritten as
\begin{equation}
A_X=
e^{-i\frac{\beta}{k''} t+i\frac{\beta^2}{2k''}z}
[\varphi_\slash(t-\beta z,r)+\varphi_\backslash(t-\beta z,r)]\text{,}
\end{equation}
with 
\begin{equation}
\varphi_X(\tau,r)=
\int_{0}^\infty e^{\mp\,i\alpha\tau}J_0(\sqrt{k''k}\alpha r)f_X(\alpha)d\alpha\text{.}
\end{equation}
The ``X'' stands for either $\slash$ or $\backslash$,
and $\varphi_X$ corresponds
to X-waves of the Helmholtz equation.\cite{Saari01}

The spatio-temporal spectrum of  $A_X(r,t,z)$ is given by the Fourier-Bessel transform pair 
\begin{equation}
\label{FourierBesselTrasform}
\begin{array}{l}
\mathcal{B}[A](k_\perp,\omega,z)=\displaystyle \frac{1}{2\pi} \int_{-\infty}^{\infty}\int_0^{\infty} 
A(r,t,z) J_0(k_\perp r) \exp(i\omega t)r dr dt\\
\\
A(r,t,z)=\displaystyle \int_{-\infty}^{\infty}\int_0^{\infty} \mathcal{B}[A](k_\perp,\omega,z) J_0(k_\perp r) \exp(-i\omega t)k_\perp dk_\perp d\omega\text{,}
\end{array}
\end{equation}
it is centered at the shifted central frequency $\beta/k''$, 
determined by the velocity:
\begin{equation}
\label{AXspectrum}
\begin{array}{l}
\displaystyle \mathcal{B}[A_X]=\displaystyle
\frac{1}{k_\perp}
\,f_\slash(\frac{k_\perp}{\sqrt{k''k}})
\delta(\omega-\frac{k_\perp}{\sqrt{k''k}}-\frac{\beta}{k''})
e^{i(\frac{\beta^2}{2k''}+\frac{\beta\,k_\perp}{\sqrt{k''k}})z}+
\\
\displaystyle \frac{1}{k_\perp}
\,f_\backslash(\frac{k_\perp}{\sqrt{k''k}})
\delta(\omega+\frac{k_\perp}{\sqrt{k''k}}-\frac{\beta}{k''})
e^{i(\frac{\beta^2}{2k''}-\frac{\beta\,k_\perp}{\sqrt{k''k}})z}\text{,}
\end{array}
\end{equation}
and it looks like an X in the angle-frequency plane.
The two terms in (\ref{AXspectrum}) correspond to slash and backslash X-waves,
as shown in figure \ref{figurepsiX}.
Since the pulsed beam travels rigidly (in modulus), an X-shaped spectrum
determines the X-shape in the $(r,t)$ space
(roughly, the far field, i.e. the Fourier-Bessel transform, resembles the near field).

The following relation holds useful for any $\psi_X(t-\beta z,r,\beta)$:
\begin{equation}
\mathcal{L}\psi_X\equiv (i\partial_z +\frac{1}{2k}\nabla^2_{xy}-\frac{k''}{2}\partial_{tt})\psi_X=-\frac{\beta^2}{2k''}\psi_X\text{.}
\end{equation}
Therefore an X-wave can also be defined as a solution of the equation 
 (\ref{NLSdimensional2}) of the type $A_X=C(z,\beta)\psi_X(t-\beta z,r)$ with
\begin{equation}
i\frac{\partial C}{\partial z}-\frac{\beta^2}{2k''}C=0\text{,}
\end{equation}
a formulation that will become handy below.

Such a solution contains, in general, an infinite energy. 
This is due to the idealized situation (never available in 
Experiments) of a precisely defined velocity (or inverse differential velocity $\beta$). 
Any finiteness introduced by the experimental setup, such as the spatial 
extension of the sample, 
will in general fade the spectrum lineshape around the X, 
determining uncertainty in $\beta$.
I will show
below that this can be described 
by a packet of X-waves, with velocities around a given value and finite energy.
Considering an X-wave with a specific velocity is thus 
as idealized as considering an elementary particle of given momentum. 
\cite{ItzyksonBook}

Before proceeding further I emphasize that entire more general classes of X-waves can be 
generated as discussed in \cite{Porras03a} or, for instance, by taking $k_z=-\beta^2/2k''-\kappa^2$ in (\ref{X1}). 
However, the case considered ($\kappa=0$) 
suffices to represent a wide class of beams, as it will be shown in the next section.
\section{X-wave expansion and finite energy solutions}
The general solution of Eq. (\ref{NLSdimensional2}) can be expressed by the Fourier-Bessel spectrum of the field at $z=0$, denoted by 
$S(\omega,k_\perp)=\mathcal{B}[A](\omega,k_\perp,0)$:

\begin{equation}
\label{generallinearsolution}
A(r,z,t)=\int_0^{\infty}\int_{-\infty}^{\infty}S(\omega,k_\perp)J_0(k_\perp r)
e^{i(k_z z-\omega t)}k_\perp dk_\perp d\omega\text{,}
\end{equation}
with $k_z=-k_\perp^2/2k+k''\omega^2/2$.

By a simple variable change, the field can be written as a superposition of slash-X-waves,
traveling with different velocities. Letting
\begin{equation}
\label{ansatzX}
\begin{array}{l}
\omega=\alpha+\frac{\beta}{k''}\\
\\
k_\perp=\sqrt{k\,k''} \alpha
\end{array}
\end{equation}
gives
\begin{equation}
A(r,z,t)=\int_{-\infty}^{\infty}
e^{-i\frac{\beta^2}{2k''}z}\Psi_\slash(t-\beta z,r,\beta)d\beta\text{,}
\end{equation}
while being
\begin{equation}
\label{Xtrasform}
\Psi_\slash(t-\beta z,r,\beta)=
\int_{0}^{\infty}X_\slash(\alpha,\beta)J_0(\sqrt{k\,k''}\alpha r)
e^{-i(\alpha+\frac{\beta}{k''})(t-\beta z)}d\alpha\text{,}
\end{equation}
and
\begin{equation} 
X_\slash(\alpha,\beta)\equiv\mathcal{X}_\slash[A(r,t,z=0)](\alpha,\beta)\equiv
k\,\alpha\,S(\alpha+\frac{\beta}{k''},\sqrt{k\,k''}\alpha)\text{.}
\end{equation}
An equivalent representation is obtained by backslash-X-waves, 
replacing the equation for $\omega$ in (\ref{ansatzX}) 
with $\omega=-\alpha+\beta/k''$, and being
\begin{equation}
X_\backslash(\alpha,\beta)=\mathcal{X}_\backslash[A(r,t,z=0)](\alpha,\beta)=
k\,\alpha\,S(-\alpha+\frac{\beta}{k''},\sqrt{k\,k''}\alpha)\text{.}
\end{equation}
The variable change (\ref{ansatzX}) corresponds
to span the $(\omega,k_\perp)$ space by oblique (slash) parallel lines in (\ref{generallinearsolution}).

Eq. (\ref{Xtrasform}) is a formulation of the ``X-wave transform'', 
first introduced in \cite{Lu00} and indicated $\mathcal{X}[A](\alpha,\beta)$.
Hence the spatio-temporal evolution, including diffraction and dispersion, 
can be represented by one dimensional propagation of 
packets with different velocities.

The energy of the pulsed beam is 
\begin{equation}
\label{energy}
\mathcal{E}=\displaystyle 2\pi\int_{0}^{\infty}\int_{-\infty}^{\infty}|A(r,t,z)|^2 r\,dr\,dt=
\int_{-\infty}^{\infty}\mathcal{E}_\beta (\beta)d\beta
\end{equation}
with
\begin{equation}
\label{energy1}
\mathcal{E}_\beta(\beta)\equiv
\displaystyle \int_0^{\infty}\frac{4\pi^2\,|X_\slash(\alpha,\beta)|^2}{k\alpha}d\alpha\text{,}
\end{equation}
showing that $\mathcal{E}_\beta$ can be taken as the energy distribution function
with respect to the inverse differential velocity $\beta$.

Summarizing, an arbitrary beam can be expanded in a superposition of X-waves
traveling with different velocities. Conversely, such a superposition can be used to
construct new-classes of physically realizable finite-energy X-waves.
To this extent, orthogonal X waves  -first introduced in \cite{Salo01b} for the 
wave equation-  are a fruitful approach (see also \cite{Conti04}).
 With reference to two (either slash
or backslash)  X-wave solutions of (\ref{NLSdimensional2}),
denoted by $A_X$ and $B_X$, with inverse differential velocities $\beta$ and $\beta'$ 
and spectra $f$ and $g$, respectively, 
the inner-product can be defined as the integral of 
$B_X^*\,A_X$ with respect to $x,y,t$, extended on the whole space:
\begin{equation}
<B_X|A_X>=\int\int\int B_X^*\,A_X\,dxdydt=\delta(\beta-\beta')
\int_0^\infty \frac{4\pi^2\,g(\alpha)^*\,f(\alpha)}{k\alpha} d\alpha \text{.}
\end{equation}
If $f=f_p$ and $g=f_q$ by defining 
($p,q=0,1,2,..$ and $L_p^{(1)}$ is the generalized Laguerre
polynomial)
\begin{equation}
f_p(\alpha)=\sqrt{\frac{k}{\pi^2(p+1)}}L_p^{(1)}(2\Delta\alpha)\Delta\alpha\,e^{-\Delta\alpha}\text{,}
\end{equation}
with $\Delta$ a parameter with the dimension of time,\cite{Conti03b} it is 
\begin{equation}
\int_0^\infty \frac{f_p(\alpha)f_q(\alpha)}{\alpha}d\alpha
=\frac{k}{4\pi^2}\delta_{pq}
\end{equation}
and the orthogonality condition ($A_q=B_X$ and $A_p=A_X$) holds:
\begin{equation}
<A_q(r,t,z,\beta)|A_p(r,t,z,\beta')>=\delta_{pq}\delta(\beta-\beta')\text{.}
\end{equation}

In the following the {\it fundamental slash-X-wave}, with spectrum 
\begin{equation}
\label{fundamentalXwavespectrum}
f_0(\alpha)=\frac{\sqrt{k}}{\pi}\Delta\alpha\exp(-\Delta\alpha)\textit{,}
\end{equation}
and spatial profile
\begin{equation}
\label{fundamentalXwaveprofile}
\varphi^{(0)}_\slash=\int_0^\infty\,f_0(\alpha)J_0(\sqrt{k\,k''}\alpha)e^{-i\alpha\,s}d\alpha=
-\frac{\sqrt{k}}{\pi}\frac{\Delta}{[1-\frac{k\,k''\,r^2}{(s-i\Delta)^2}]^{3/2}(s-i\Delta)^2}
\text{,}
\end{equation}
will be taken as a prototype of the simplest X-wave with finite power
(i.e. converging transverse integral of $|\varphi^{(0)}_\slash|^2$, see also \cite{Ciattoni04}).
Its intensity profile can be found, e.g., in \cite{Conti04}.

When $X(\alpha,\beta)=C(\beta)f_p(\alpha)$, it is $\mathcal{E}_\beta=|C(\beta)|^2$, and
\begin{equation}
\mathcal{E}=\int_{-\infty}^{\infty}|C(\beta)|^2\,d\beta\text{.}
\end{equation}
The resulting beam is a finite energy X-wave that spreads
according to a prescribed velocity distribution function $C(\beta)$;
this corresponds to the existence of solutions with an arbitrary ``depth of focus''.

This kind of wave-packet can be written, with reference to slash X-waves, as
\begin{equation}
A=\int_{-\infty}^{\infty}C(\beta,z)\psi_\slash^{(q)}(r,t-\beta\,z,\beta)d\beta,
\end{equation}
(with obvious notation) while being, as above,
\begin{equation}
\label{largeceq}
i\frac{\partial\,C}{\partial\,z}-\frac{\beta^2}{2k''}C=0\text{.}
\end{equation}
By introducing the Fourier transform pair of $C(\beta)$
\begin{equation}
\label{smallcdef}
\begin{array}{l}
c(s,z)=\frac{1}{2\pi\sqrt{k''}}
\int_{-\infty}^{\infty}C(\beta,z)e^{i\frac{\beta}{k''}s}d\beta\\
\\
C(\beta,z)=\frac{1}{\sqrt{k''}}\int_{-\infty}^{\infty}
c(s)e^{-i\frac{\beta}{k''}s}ds
\text{,}
\end{array}
\end{equation}
with $s$ in time units ($s$ can be roughly kept in mind as the on-axis temporal variable $t$), it is from (\ref{largeceq}) 
\begin{equation}
i\frac{\partial\,c}{\partial\,z}+\frac{k''}{2}\frac{\partial^2\,c}{\partial\,s^2}=0\text{.}
\end{equation}
Hence 3D linear propagation of an X-wave packet in a normally dispersive medium
can be reduced to that of a 1D pulse with anomalous dispersion.
The energy is 
\begin{equation}
\mathcal{E}=2\pi\int_{-\infty}^{\infty}|c(s)|^2\,ds\text{.}
\end{equation}

As discussed in \cite{Conti04}, by expanding $X(\alpha,\beta)$ into generalized Laguerre polynomials
it is possible to express the general solution of (\ref{NLSdimensional2})
by orthogonal X-waves (either $\slash$ or $\backslash$).
This has been used in \cite{Conti04b} as an approach to field quantization, and applied 
to quantum optical parametric amplification.
\section{General properties of X-wave propagating in the nonlinear regime} 
Since X-waves exist even in the absence of a nonlinear susceptibility, it is legitimate to resort to 
a perturbative approach. A basic model for nonlinear optical interactions (under standard approximations) 
can be cast in the form
\begin{equation}
\label{generalnonlinear}
i\partial_z A+\frac{1}{2k}\nabla^2_{xy}A-\frac{k''}{2}\partial_{tt}A=\chi\mathcal{P}_{NL}(z,t,r) \text{,}
\end{equation}
where $\mathcal{P}_{NL}(z,t,r)$ is a nonlinear source, weighed by $\chi$. 
After a straightforward expansion of $A$ in powers of $\chi$, the relevant equation
is (\ref{generalnonlinear}) at any order, where 
the right-hand side stems from the solutions at lower orders.
I will show below that if 
$\chi\mathcal{P}_{NL}(z,t,r)=\mathcal{P}(t-\bar{\beta}z,r)$,
the evolution according to (\ref{generalnonlinear}) always provides a spatio-temporal spectrum corresponding
to a progressive undistorted wave. Thus, if an X-wave is taken as the solution of the linear model ($\chi=0$), 
the nonlinearity has the role of ``dressing'' that
solution, which still continues to exist. 
Since this result is valid at any order, it furnishes a general picture of the propagation 
of self-invariant beams in the nonlinear regime. 

For an X-wave propagating in a nonlinear medium, $\mathcal{P}$
can be interpreted as a function of its field and of its complex conjugate. 
In the case of harmonic generation, $\mathcal{P}$ is some power of the pump beam,
traveling at the group velocity of the fundamental frequency.
\cite{noteA2}
What follows can be viewed as a generalization of what discussed in  \cite{Conti03b},
with the inclusion of second order dispersion and for a wide class of nonlinear processes
(such as third and higher harmonics generation).

In a Kerr medium, with a refractive index $n=n_0+n_2\,I$, with
$I=|A|^2$ the optical intensity, Eq. (\ref{generalnonlinear}) becomes
\begin{equation}
\label{NLSnonlinear}
i\partial_z A+\frac{1}{2k}\nabla^2_{xy}A-\frac{k''}{2}\partial_{tt}A+
\frac{k\,n_2}{n_0}\,I\,A=0\text{.}
\end{equation}
For the solution at the lowest order ($n_2=0$), it is possible to take either an X-wave 
packet $A_X$ around $\bar\beta$ or a wide pulsed beam with negligible diffraction 
(with $\bar\beta=0$). Higher orders are obtained in the form 
(\ref{generalnonlinear}):  at the first the rhs is proportional to $A_X^2\,A_X^*$.
 As a result of the following analysis, 
the correction to $A_X$ is still a progressive undistorted wave
traveling at the same velocity. Since this can be 
applied at any order, linearly-self-invariant beams are
very robust with respect to the nonlinearity. 

By writing $A$, as a superposition of slash-X-waves,
\begin{equation}
\label{super1}
A=\int_{0}^{\infty} \int_{-\infty}^{\infty}  C(\alpha_1,\beta_1,z) J_0(\sqrt{kk''}\alpha_1 r) e^{-i(\alpha_1+\beta_1/k'')(t-\beta_1 z)}d\alpha_1\,d\beta_1\text{,}
\end{equation}
and inserting into (\ref{generalnonlinear}), one obtains
\begin{equation}
\label{tempor1}
\int_{-\infty}^{\infty} \int_{0}^{\infty}  
(i\frac{\partial C}{\partial z}-\frac{\beta_1^2}{2k''}C )J_0(\sqrt{kk''}\alpha_1 r) 
e^{-i(\alpha_1+\beta_1/k'')(t-\beta_1 z)}d\alpha_1\,d\beta_1
=\mathcal{P}(t-\bar{\beta}z,r)\text{.}
\end{equation}
Taking the Fourier-Bessel transform of Eq.(\ref{tempor1}), 
multiplying it by $2\pi r J_0(k_\perp r)\exp(i\omega t)$ 
and integrating over $r$ and $t$, one founds
\begin{equation}
\label{Cequation}
i\frac{\partial C}{\partial z}-\frac{\beta^2}{2k''}C =
\mathcal{X}_\slash[P](\alpha,\beta)
e^{i(\alpha+\beta/k'')(\bar\beta-\beta)z}\text{,}
\end{equation}
where Eqs. (\ref{ansatzX}) apply, and
\begin{equation}
\mathcal{X}_\slash[P](\alpha,\beta)=
k\alpha\mathcal{B}[P](\alpha+\frac{\beta}{k''},\sqrt{k\,k''}\alpha)\text{.}
\end{equation}

Eq. (\ref{Cequation}) can be readily integrated with the boundary condition $C=0$ at $z=0$:
\begin{equation}
\label{Csolution}
C(\alpha,\beta,z)=-i\mathcal{X}_\slash[P]
\frac{sin(g z)}{g}e^{i[(\alpha+\frac{\beta}{k''})
(\bar\beta-\beta)-\frac{\beta^2}{4k''}]z}\text{,}
\end{equation}
with 
\begin{equation}
g=\frac{1}{2}[(\alpha+\frac{\beta}{k''})(\bar\beta-\beta)+\frac{\beta^2}{2k''}]\text{.}
\end{equation}
Eq. (\ref{Csolution}) can be interpreted in the $(\omega,k_\perp)$, or $(\alpha,\beta)$, 
planes and shows that, for large propagation distances, $C$ tends 
to a Dirac $\delta$ centered at $g=0$: 
the propagation acts as a spatio-temporal filter, selecting specific combinations of frequencies and wavevectors.

\begin{figure}
\includegraphics[width=80mm]{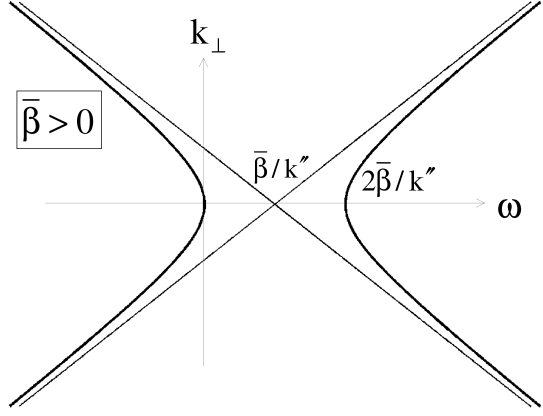}
\caption{\label{figureiperbola} 
Sketch of the generated spatio-temporal spectrum
 (symmetrized for $k_\perp<0$) when $\bar\beta>0$.
The straight lines are slash and backslash spectra.}
\end{figure}

The condition $g=0$, in the $(\omega,k_\perp)$ plane, is
\begin{equation}
\label{iperbola}
\frac{k''}{2}(\omega-\frac{\bar\beta}{k''})^2-\frac{k_\perp^2}{2k}=\frac{\bar\beta^2}{2k''} 
\end{equation}
which gives a hyperbola, as sketched in figure \ref{figureiperbola}.
It is apparent that Eq. (\ref{iperbola}) is the dispersion relation corresponding
to Eq. (\ref{X1}), therefore the resulting pulsed beam is a progressive undistorted wave traveling
at inverse differential velocity $\bar\beta$. 
In reported experimental results there is a clear evidence of this spectral hyperbola, see e.g. \cite{DiTrapaniPRFocus},
as well as of a structured spectrum,\cite{Nibbering96,Faccio2004} as that corresponding to the splitting described in the following. 

The asymptotes over which energy is concentrated 
correspond to slash and backslash X-waves.
Thus, in any nonlinear process which can be
reduced to equation (\ref{generalnonlinear}), X-wave packets are spontaneously generated. 
I like to stress that one of the arms always starts at $\omega=0$, while in the case $\bar\beta=0$
 Eq. (\ref{iperbola}) yields the exact X-shaped spectrum.
The specific features will in general depend on the source spectrum
, as elucidated below with an example.

Note also that the same analysis holds true, with simple changes, when 
$\chi\mathcal{P}_{NL}(z,t,r)=\mathcal{P}(t-\bar{\beta}z,r)\exp(i\bar k_z z)$, with $\bar k_z$ depending on the specific
nonlinear process (e.g. $\bar k_z$ is the wavevector mismatch for optical harmonics generation), detailed cases will
be reported elsewhere.
\subsection{Energy of the generated X-wave}
The energy distribution function is, after (\ref{energy1}), 
\begin{equation}
\mathcal{E}_\beta(\beta)=\int_0^{\infty}
\frac{4\pi^2|C(\alpha,\beta,z)|^2}{k\alpha}d\alpha=
\int_0^{\infty}
\frac{4\pi^2 |\mathcal{X}_\slash[P]|^2\,(\alpha,\beta)}{k\alpha}\frac{sin(gz)^2}{g^2}\,d\alpha\text{.}
\end{equation}
As $z\rightarrow\infty$
\begin{equation}
\frac{sin(gz)^2}{g^2}\rightarrow\frac{2\pi\,z}{|\bar\beta-\beta|}\delta(\alpha-\bar\alpha)
\text{,}
\end{equation}
with
\begin{equation}
\label{alfabar}
\bar\alpha\equiv\frac{\beta(\beta-2\bar\beta)}{2k''\,(\bar\beta-\beta)}\text{.}
\end{equation}
Hence for large $z$ 
\begin{equation}
\label{asymptenergy}
\mathcal{E}_\beta=\frac{8\pi^3 k\bar\alpha\,z}{|\beta-\bar\beta|}
 |\mathcal{B}[P](\bar\alpha+\frac{\beta}{k''},\sqrt{kk''}\bar\alpha)|^2
\theta_0(\bar\alpha)\text{,}
\end{equation}
with $\theta_0$ the unit step function.

The generated X-wave packet grows with a linear efficiency 
with respect to the propagation distance.
In the case $\bar\beta\neq0$, from the condition $\bar\alpha>0$
in (\ref{asymptenergy}) it appears that energy is distributed
in the intervals $\beta<0$ and $\bar\beta<\beta<2\bar\beta$ when $\bar\beta>0$, 
and in $\bar\beta<\beta<0$ and $\beta<2\bar\beta$ when $\bar\beta<0$.
This implies that $\mathcal{E}_\beta$ does not have a continuous support (see the example in figure \ref{figurebeta}, discussed below)
For large $|\bar\beta|$, with sufficiently separated domains along $\beta$, 
this is expected to provide pulse-splitting.
In fact, some components of the generated X-wave will travel at a velocity sensibly different from
the pump wave, and after some propagation length satellite packets may appear (such as, e.g., those discussed in \cite{Kolesik04}).

There are two relevant limits for Eq. (\ref{asymptenergy}).

As $\bar\beta\rightarrow0$, i.e. the pump travels at the linear group velocity of the medium,
and $\bar\alpha\rightarrow-\beta/2k''$, it is 
\begin{equation}
\label{asymptenergy0}
\mathcal{E}_\beta\rightarrow \frac{4\pi^3 k\,z}{\,k''}
 |\mathcal{B}[P](\frac{\beta}{2k''},-\sqrt{kk''}\frac{\beta}{2k''})|^2
\theta_0(-\beta)\text{.}
\end{equation}
Eq. (\ref{asymptenergy0}) shows that only the portion of the pump spectrum 
laying on the X-spectrum, i.e. on the line $k_\perp=-\sqrt{k\,k''}\omega$, contributes
to the energy of the generated beam. The propagation filters out 
all of the spectral components that do not belong to an X-wave.

The other relevant limit corresponds to a pump velocity largely different from
the linear group velocity of the medium: $|\bar\beta|\rightarrow\infty$.
From a physical point of view, the nonlinear interaction is 
strongly hampered by the generated beam which rapidly gets separated from 
the pump, producing a typical ``Cerenkov emission'' halo. 
From (\ref{asymptenergy}) it is ($\bar\alpha\rightarrow-\beta/k''$)
\begin{equation}
\label{asymptenergyinfty}
\mathcal{E}_\beta\rightarrow \frac{8\pi^3  k\,|\beta|\,z}{\,k''|\bar\beta|}
 |\mathcal{B}[P](0,-\sqrt{kk''}\frac{\beta}{k''})|^2
\theta_0(-\beta)\text{.}
\end{equation}
For large $|\bar\beta|$ the X-wave generation is strongly
inhibited, and exclusively the on-axis spatio-temporal spectrum plays a role.
Only the branch $\beta<0$
is relevant in this limit. Furthermore, because of the factor $|\beta|$, the energy distribution is expected to be peaked at some value $\beta<0$, as shown
below. 
\subsection{Example: Gaussian pump beam}
It is instructive to examine an example, by assuming a Gaussian lineshape for the source:
\begin{equation}
\label{pumpexample}
\mathcal{B}[P](k_\perp,\omega)=\mathcal{B}_0^2\,exp(-\frac{k_\perp^2}{\Delta k_P^2}-\frac{\omega^2}{\Delta\omega_P^2})\text{,}
\end{equation}
such that $\mathcal{B}_0^2$ measures the pump beam fluence, and $\Delta\omega_P$ and $\Delta k_P$ provide
 the on axis temporal and spatial spectra, respectively. 

In the case $\bar\beta=0$ (source traveling at the linear group velocity at $\omega_0$) 
the energy distribution function, after Eq. (\ref{asymptenergy}), is 
\begin{equation}
\mathcal{E}_\beta=\frac{4\pi^3 \mathcal{B}_0^2\,k\,z}{k''}
\exp(-\frac{\beta^2}{\Delta\beta_P^2})\theta_0(-\beta)
\end{equation}
showing that the generated spectrum encompasses slash X-waves faster than the source
($\beta<0$), and spread in velocity with the characteristic value 
\begin{equation}
\Delta\beta_P^2=[\frac{1}{4(k'')^2\,\Delta\omega_P^2}+\frac{k}{4\,k''\,\Delta k_P^2}]^{-1}\text{,}
\end{equation}
such that the more the source is spectrally (in space or time)
narrow the more ``ideal'' (i.e. un-spreading, $\Delta\beta_P$ small) is the generated X-wave-packet.

In the case $\bar\beta\neq0$ the situation is more complicated. 
From (\ref{asymptenergy}) and (\ref{pumpexample}),  it is
$\mathcal{E}_\beta=0$ when $\bar\alpha(\beta)<0$; while for other values of $\beta$
\begin{equation}
\label{energybeta}
\mathcal{E}_\beta=
\frac{4\pi^3 \mathcal{B}_0^2\,k\,z}{k''}
\frac{\beta^2-2\beta\bar\beta}{(\beta-\bar\beta)^2}
\exp[-\frac{\beta^2}{(\beta-\bar\beta)^2}
\frac{\beta^2-q_P(\beta\bar\beta-\bar\beta^2)}{\Delta\beta_P^2}]
\end{equation}
with $q_P=k\,\Delta\beta_P^2/k''\Delta k_P^2$.

\begin{figure}
\includegraphics[width=120mm]{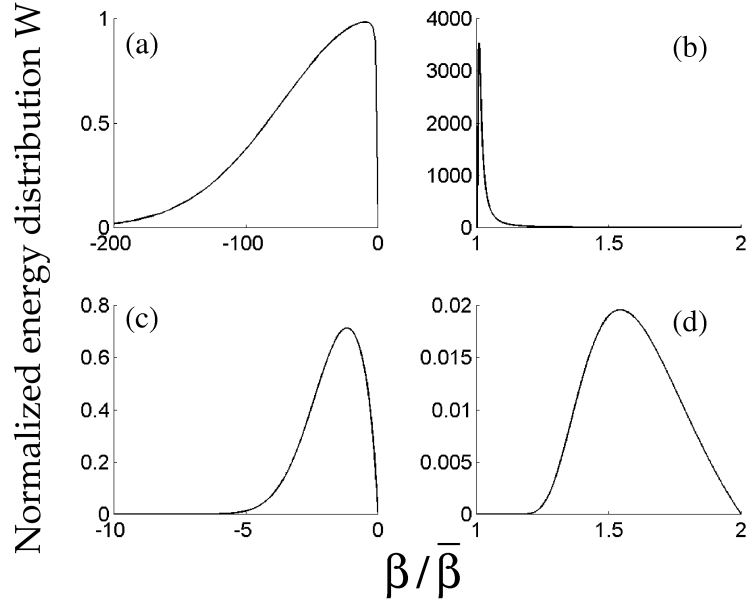}
\caption{\label{figurebeta}  
Normalized energy distribution $W$ Vs inverse differential velocity $\beta$ for two different 
values of pump velocity determined by $\bar\beta$. 
(a)  branch $\beta<0$  for $\bar\beta=0.01\Delta\beta_P$;
(b)  branch $\bar\beta<\beta<2\Delta\bar\beta_P$ for $\bar\beta=0.01\Delta\beta_P$;
(c) as in (a) with $\bar\beta=0.5\Delta\beta_P$;
(d) as in (b) with $\bar\beta=0.5\Delta\beta_P$.
All quantities are dimension-less.
}
\end{figure}

In figure \ref{figurebeta} I show an example of the application of Eq. (\ref{energybeta}),
 by reporting the normalized energy distribution $W=\mathcal{E}_\beta k''/4\pi^3 \mathcal{B}_0^2 k z$.
The pump beam, with $\omega_0$ such that the wavelength is $\lambda_0=1\mu\,m$, 
is chosen with $\Delta\omega_P=2\pi/\Delta t$, 
with $\Delta t=500fs$, and $\Delta k_P=\Delta \theta k_0$ with $\Delta \theta=0.1 rad$.
For the medium parameters $n_0=2$, $k'\cong c/n_0$, $k''=360\times10^{-28}s^2/m$.
As a result, $\beta_p\cong10^{-12}s/m$, with a velocity bandwidth 
$\cong\Delta\beta_P\,V_g^2\cong10^4\,m/s$, and $q_p\cong10^{-4}$.

In figure \ref{figurebeta},  (a) and (b), 
the two branches are displayed for $\bar\beta=0.01\Delta\beta_P$,
corresponding to a pump beam velocity of $V_P=0.499999c$.
In this case, the energy distribution tends to that at $\bar\beta=0$, and
no pulse splitting is expected (the two branches are almost contiguous).

In figure \ref{figurebeta}, (c) and (d), 
the two branches are shown for $\bar\beta=0.5\Delta\beta_P$,
corresponding to a pump beam velocity of $V_P=0.499966c$. Two distinct lobes
are present, with the most energetic peaked around $\beta\cong-2\bar\beta$.
In this case a satellite X-wave packet is expected to separate from the 
pump beam in propagation.

The X-wave will in general split for sufficiently long propagation, following the
peaks of $\mathcal{E}_\beta$. 
At higher orders the resulting packets will become new sources
for the perturbation equation, and additional splitting will appear, consistently
with the numerical results (see e.g. \cite{Berge02,Kolesik04}).

This is expected in second-harmonic generation, 
initially considered in \cite{Conti03b}, where $\bar\beta$ is the temporal walk-off
between the two harmonics (details will be given elsewhere);
 as well as in dispersive 
propagation, as described in \cite{Kolesik04}, 
and investigated experimentally in \cite{DiTrapani03b}.
 The spectral widening can indeed
provide sources traveling at a group velocity different from that at $\omega_0$
($1/k'$).

In figure \ref{figurebeta} note the difference in the vertical scale between 
 panels (a), (b) and (c), (d). 
In all cases the pump velocity is about $1/k'$, showing that the mismatch 
between the pump and the linear group velocities at $\omega_0$
drastically affects the generated X-wave spectrum.

With regards to the efficiency of X-wave generation, in the case $\bar\beta=0$, 
by integrating 
(\ref{asymptenergy0}) it is found 
\begin{equation}
\label{efficiency0}
\mathcal{E}=\frac{2\pi^{7/2}\mathcal{B}_0^2\,k\,z}{k''}\Delta\beta_P=
4\pi^{7/2} \mathcal{B}_0^2\,k\,z\frac{\Delta\omega_P}{\sqrt{1+
\frac{k''\,k\,\Delta\omega_P^2}{\Delta k_P^2}}}\text{.}
\end{equation}
The energy grows with $\Delta\beta_P$, i.e. with an extension 
of the spatio-temporal spectrum of the pump. 
After Eq.(\ref{efficiency0}) one founds that, for a wide spatial spectrum (i.e. large $\Delta k_P$,
corresponding to a tightly focused pump beam), the efficiency is mainly determined by the on-axis temporal spectrum.
Conversely, for large on-axis spectrum $\Delta\omega_P$, the energy
of the generated X-wave grows with $\Delta k_P$.

The other relevant limit is $\bar\beta>>\Delta\beta_P$, such that the resulting 
energy distribution is solely determined by the branch $\beta<0$, and it is 
\begin{equation}
\label{efficiency1}
\mathcal{E}=\frac{2\pi^3 \mathcal{B}_0^2\,z\,\Delta k_P^2}{\bar\beta}\text{.}
\end{equation}
Thus the efficiency is mainly affected by the spatial profile of the pump 
beam and it goes like $1/|\bar\beta|$, consistently with Eq. (\ref{asymptenergyinfty}).
\section{Highly nonlinear regime: an effective 1D NLS}
Even when using some working hypotheses, it is possible to deepen the previous study 
and consider the highly nonlinear regime.
I will consider the
multidimensional Schroedinger equation:
\begin{equation}
\label{NLSdimensional3}
i\partial_Z A+ik'\partial_T A+\frac{1}{2k}\nabla^2_{xy}A-\frac{k''}{2}\partial_{TT}A+\frac{k n_2}{n_0}|A|^2 A=0\text{.}
\end{equation}
This may provide the correct trend observed in the experiments, including
quadratic nonlinearity in certain regimes, (see e.g.\cite{Torruellas01}) as long as 
envelope shocks (see \cite{Berge02}), or higher order phenomena such as plasma
formation play a negligible role. 

Writing $A$ as a  X-wave expansion ($0<\alpha<\infty$ and $-\infty<\beta<\infty$):
\begin{equation}
A=\int X(\alpha,\beta,z)J_0(\sqrt{k\,k''}\alpha r)e^{-i(\alpha+\beta/k'')(t-\beta z)} d\alpha d\beta\
\end{equation}
 and taking the Fourier Bessel transform of Eq. (\ref{NLSdimensional3}), 
evaluated at $k_\perp=\sqrt{k k''}\alpha$,
and $\omega=\alpha+\beta/k''$,
the {\it coupled X-wave equations} are derived:
\begin{equation}
\label{coupledXwaves}
\begin{array}{l}
i\partial_z X(\alpha,\beta,z)-\frac{\beta^2}{2k''}X(\alpha,\beta,z)
+\frac{k\,n_2}{n_0}\mathcal{Q}(\alpha,\beta,z)=0\text{.}
\end{array}
\end{equation}
The nonlinear polarization is 

\begin{equation}
\label{NLSQ}
\begin{array}{l}
\mathcal{Q}=\mathcal{X}_\slash[|A|^2\,A^*]e^{-i(\alpha+\frac{\beta}{k''})\beta\,z}
=\\
\\
k\alpha\int\,\mathcal{K}(\alpha,\vec\alpha)
X(\alpha_1,\beta_1,z)\,X(\alpha_2,\beta_2,z)\,X(\alpha_3,\beta_3,z)^* \\
\\
\delta[\alpha+\alpha_3-\alpha_1-\alpha_2+(\beta+\beta_3-\beta_1-\beta_2)/k'']
\Theta(\alpha,\beta,\vec\alpha,\vec\beta,z)d\vec\alpha d\vec\beta\text{,}
\end{array}
\end{equation}
with $\vec{\alpha}=(\alpha_1,\alpha_2,\alpha_3)$,
$\vec{\beta}=(\beta_1,\beta_2,\beta_3)$,\cite{SpecialFunctionsBook}
\begin{equation}
\mathcal{K}(\alpha,\alpha_1,\alpha_2,\alpha_3)
=\int_0^\infty J_0(\sqrt{k k''}\alpha r) J_0(\sqrt{k k''}\alpha_1 r)
J_0(\sqrt{k k''}\alpha_2 r) J_0(\sqrt{k k''}\alpha_3 r )r\,dr
\end{equation}
and
\begin{equation}
\Theta=\exp\{i[-(\alpha+\beta/k'')\beta+(\alpha_1+\beta_1/k'')\beta_1+
(\alpha_2+\beta_2/k'')\beta_2-(\alpha_3+\beta_3/k'')\beta_3]z\}\text{.}
\end{equation}

A remarkable result is obtained if the solution of (\ref{coupledXwaves}) is approximated by 
an X-wave-packet which, in order to avoid unnecessary complexities,
is centered around the medium group
velocity, i.e. $\bar\beta=0$. This is standard 
in usual coupled mode theory; \cite{MoloneyBook} in the present case
a ``mode'' is an X-shaped 3D wavepacket.

As discussed below, if $z$ is sufficiently small, it is possible to take
$\Theta\cong1$ in (\ref{NLSQ}); in this regime I write
 $X(\alpha,\beta)=f(\alpha)C(\beta)$ in order to represent the 
nonlinear modulation (with envelope $C$) of an X-wave. 
For definiteness, I consider $f=f_p$,
where $f_p(\alpha)$ is 
the spectrum of a basis X-wave.
If $C$ is peaked around $\bar\beta=0$, all components
 travel nearly at the same velocity and $\Theta\cong1$.
Multiplying by $4\pi^2\,f_p(\alpha)/k\,\alpha$ and integrating over $\alpha$,
from Eq. (\ref{coupledXwaves}) one obtains:

\begin{equation}
\label{spectralXNLS}
\begin{array}{l}
i\partial_z C(\beta,z)-\frac{\beta^2}{2k''}C(\beta,z)
+\frac{k\,n_2}{n_0}\int \chi(\beta+\beta_1-\beta_2-\beta_3)  
C(\beta_1)\,C(\beta_2)\,C(\beta_3)^*
d\vec\beta=0  \text{.}
\end{array}
\end{equation}
with the interaction kernel:
\begin{equation}
\chi(\gamma)=4\pi^2\int  
\,\mathcal{K}(\alpha,\alpha_1,\alpha_2,\alpha_3)\\
f(\alpha) f(\alpha_1)\,f(\alpha_2)^*\,f(\alpha_3)^*\\
\delta[\alpha+\alpha_1-\alpha_2-\alpha_3+\frac{\gamma}{k''}]\\
d\vec\alpha\text{.}
\end{equation}

Eq. (\ref{spectralXNLS}) is a Zakharov equation,
and taking the Fourier transform of $C$ (see (\ref{smallcdef})) one obtains 
the 1D nonlinear Schroedinger equation (NLS):
\begin{equation}
\label{1DNLS}
i\frac{\partial\,c}{\partial\,z}+\frac{k''}{2}\frac{\partial^2\,c}{\partial\,s^2}
+\frac{k\,n_2}{n_0}\sigma(s)|c|^2\,c=0\text{.}
\end{equation}
Hence {\it the evolution of an X-wave packet in a nonlinear Kerr medium can be approximated by 
an effective 1+1D nonlinear Schroedinger equation with a non-homogeneous nonlinearity.}
The latter, given by $\sigma(s)$ with dimensions of an inverse area,
is expressed by the Fourier transform of the interaction kernel $\chi$:
\begin{equation}
\sigma(s)=4\pi^2\,k''\,\int_{-\infty}^{\infty}\chi(\gamma)e^{i\gamma\,s/k''}d\gamma\text{.}
\end{equation}
After some algebra it can be written as 
\begin{equation}
\sigma(s)=\int_0^{\infty}|2\pi\sqrt{k''}\varphi_\slash^{(p)}|^4 r\,dr\text{,}
\end{equation}
which is the spatial overlap of the component X-wave profile at $\beta=0$, with
\begin{equation}
\varphi_\slash^{(p)}=\varphi_\slash^{(p)}(r,s)=\int_0^{\infty}f_p(\alpha)
J_0(\sqrt{k\,k''}\alpha\,r)e^{-i\alpha\,s}d\alpha\text{.}
\end{equation}

Taking for $c$ a solution of (\ref{1DNLS}), and $C$ its Fourier transform 
according to (\ref{smallcdef}), $A$ reads
\begin{equation}
\label{AVsc}
A=\int C(\beta,z)\psi_\slash^{(p)}(r,t-\beta z,\beta)d\beta=
\int c(s,z)\xi_\slash^{(p)}(r,t,s,z)ds\text{,}
\end{equation}
with
\begin{equation}
\label{waveletbasis}
\xi_\slash^{(p)}(r,t,s,z)=\frac{1}{\sqrt{k''}}
\int e^{-i(s+t)\frac{\beta}{k''}+i\frac{\beta^2}{k''}z}
\varphi_\slash^{(p)}(r,t-\beta z)d\beta\text{.}
\end{equation}

$\xi_\slash^{(p)}$ can be considered the basis for the wavelet transform of A with respect
to the on-axis temporal variable.\cite{Keiser92} If $\Theta\cong1$ the spreading due to 
the different velocities of the component X-waves 
(given by the quadratic terms in $\beta$ in the exponential in (\ref{waveletbasis}))
 is negligible, and $\xi$ can
be approximated by its value at $z=0$:
\begin{equation}
\xi_\slash^{(p)}(r,t,s,z)\cong\xi_\slash^{(p)}(r,t,s,0)=2\pi\sqrt{k''}\delta(t+s)\varphi_p^{(p)}(r,t)
\end{equation}
and (\ref{AVsc}) becomes 

\begin{equation}
\label{AVscSemplified}
A\cong2\pi\sqrt{k''}c(-t,z)\varphi_\slash^{(p)}(r,t)\text{.}
\end{equation}

The 3D+1 nonlinear evolution problem (\ref{NLSdimensional3}) reduces, under suitable approximations, to 
a 1+1D model (\ref{1DNLS}). In the following I will address the hypotheses leading to this result and to its straightforward consequences.

\section{Conditions for observing solitons}
Eq. (\ref{1DNLS}) is valid as far as $\Theta\cong1$ in (\ref{NLSQ}).
Writing
\begin{equation}
\Theta=exp(i\theta\,z)\text{,}
\end{equation}
with
\begin{equation}
\theta=(\alpha_1+\beta_1/k'')\beta_1+(\alpha_2+\beta_2/k'')\beta_2-
(\alpha+\beta/k'')\beta-(\alpha_3+\beta_3/k'')\beta_3\text{,}
\end{equation}
it must be $|\theta|z<<1$.

$\theta$ is a quadratic function of $\beta$s;
if $\delta\beta>0$ is the velocity bandwidth ($\beta<|\delta\beta|$), 
and observing that the spectrum
of the fundamental X-waves decays as $exp(-\alpha\Delta)$ 
(thus, roughly, $\alpha<1/\Delta$), 
 the validity of this approximation leads to a constrained maximization of $\theta$.
After a simple analysis, it must be
\begin{equation}
z<<\frac{1}{\max(|\theta|)}=z_X
\end{equation}
while, if $1/\Delta<2\delta\beta/k''$, \textit{the soliton regime}:
\begin{equation}
\label{conditionX1}
\frac{1}{z_X}=2\delta\beta(\frac{1}{\Delta}+\frac{\delta\beta}{k''})+
\frac{k''}{2\Delta^2}\text{;}
\end{equation}
and, if $1/\Delta>2\delta\beta/k''$, \textit{the chirp regime}:
\begin{equation}
\label{conditionX2}
\frac{1}{z_X}=\frac{4\delta\beta}{\Delta}\text{.}
\end{equation}
As expected, the smaller $\delta\beta$, the larger $z_X$.

The above can be recast in a much simpler and more insightful formulation by 
introducing the dispersion ($L_{disp}$) and the diffraction 
($L_{diff}$) lengths. 

If $\mathcal{T}_0=k''/2\delta\beta$ is the s-width of $c$, which in general does not
 correspond to
the on-axis temporal duration of the pulsed beam (this holds approximately
only in the soliton regime, see (\ref{AVscSemplified})), it is 
$L_{disp}=\mathcal{T}_0^2/k''$.

If $\mathcal{W}_0=1/\Delta k_\perp$ is the beam waist at the center of the pulse, 
with $\Delta k_\perp$ the spatial spectrum, by the general 
properties of X-waves it is ($\alpha<1/\Delta$) 
$\mathcal{W}_0=\Delta/\sqrt{k\,k''}$, and $L_{diff}=k\,\mathcal{W}_0^2$.

When $\Delta$ is much smaller than $\mathcal{T}_0$, the X-wave dominates the 
whole spatio-temporal beam profile, this is \textit{the chirp regime}.
Conversely, when $\mathcal{T}_0<\Delta$, 
\textit{the soliton regime} is attained. In this case, the on-axis temporal
duration is also affected by the envelope $c$, while the X-wave mainly
determines the spatial profile (see (\ref{AVscSemplified})).

For instance, assuming to be in the soliton regime
(as witnessed by non-trivial nonlinear dynamics),
taking a beam waist of $\mathcal{W}_0=50\mu\,m$ and an on-axis temporal duration 
$\mathcal{T}_0=200fs$ it
is $z_X\cong8cm$, taking $n_0=2$, $k''=360\times\,10^{-28}s^2/m$ and $\lambda_0=500nm$, 
a value far beyond the propagation distances reported in 
experiments. \cite{DiTrapani03,DiTrapani03b} $z_X$ is typically of the order 
of the smallest between $L_{disp}$ and $L_{diff}$, because
it measures the distance after which the X-wave packet starts to spread due to
its finite energy. 
Therefore one expects the approximations leading to the 1D NLS model Eq.(\ref{1DNLS})
to remain valid at least in the early stages of the dynamics, a few centimeters in experiments. 
When $z$ increases, the nonlinear response gets reduced 
because of the sliding between the component X-waves, and the 
propagation becomes in essence the linear evolution of the generated X-wave pattern.

Once the validity of (\ref{1DNLS}) is ascertained,
the corresponding nonlinear dynamics is determined by considering the
the \textit{effective nonlinear length} $L_{NL}=n_0/(k\,n_2\,\sigma(0)\,c_0^2)$,
with $c_0=|c(s=0,z=0)|^2$, which relates to the input 
peak intensity, omitting inessential numerical factors,
 by $c_0^2=\mathcal{I}_0\,\mathcal{W}_0^2$
(i.e. $c_0^2$ is approximately given by the peak power, see below).
In the definition of $L_{NL}$, the
value of $\sigma(s)$ is considered at $s=0$, because prominent nonlinear effects are expected near the peak of the pulsed beam. 

The depth of focus
of the progressive undistorted wave is such that the on-axis intensity 
keeps constant in propagation, thus enhancing nonlinear effects.
This is quite similar to the nonlinear dynamics of guided modes, where 
the tightly confined light may favor nonlinear phenomena.

When the temporal duration $\mathcal{T}_0$ of the envelope $c$ is
smaller than the duration of the modulated X-wave, approximately 
$\Delta$, 
as far as $L_{disp}>>L_{NL}$ the nonlinearity will mainly play as a self phase modulation. When $L_{disp}\cong\,L_{NL}$ the so called
$N=1$ fundamental soliton solution (with a sech profile for $c$, and
$N$ the number of eigenvalues in the inverse scattering problem of ($\ref{1DNLS}$), with
$\sigma(s)\rightarrow \sigma(0)$) is attained. 
However, it can be readily seen that $L_{disp}>z_X$ 
(in the previous example $L_{disp}\cong1m$), 
hence the X-wave dynamics dominates the dispersion.
Even if a $N=1$ soliton emerges, it essentially behaves as a non-dispersing 
non-diffracting X-wave. 

The situation can be completely different when the peak power  increases and
$L_{NL}$ reduces. In this case the parameter $N$, 
given by $\sqrt{L_{disp}/L_{NL}}$,
becomes greater than unity, and hence higher-order solitons and breathers are foreseen.
They will be discussed below.
\section{Fundamental X-wave/Fundamental soliton}
Under suitable conditions, satisfied in the early stages of propagation, the nonlinear dynamics
of an X-wave in a Kerr medium can be described by the 1D nonlinear Schroedinger 
equation with a non-uniform nonlinearity profile $\sigma(s)$, 
determined by the spatial (self-)overlap of the progressive undistorted wave.

As I have shown earlier, the whole X-wave space is spanned by the fundamental X-waves.
The fundamental X-shaped profile, integrable with respect to $r$ and typically observed
in simulations and experiments, is given by $f_0(\alpha)=(\sqrt{k}/\pi)\Delta\alpha\exp(-\Delta\alpha)$ 
(see (\ref{fundamentalXwavespectrum})). 
I will consider the effective 1D NLS with such spectrum. 
It can be found that
\begin{equation}
|\varphi_\slash^{(0)}|^4=\frac{k^2\Delta^4}{\pi^4}
\frac{(s^2+\Delta^2)^2}
{[(s^2+\Delta^2)^2+(k\,k''\,r^2)^2-2(s^2-\Delta^2)k\,k''\,r^2]^3}\text{,}
\end{equation}
which gives
\begin{equation}
\sigma(s)=\frac{8k''\,k}{5\Delta^2}f_\sigma(\frac{s}{\Delta})\text{.}
\end{equation}
\begin{figure}
\includegraphics[width=80mm]{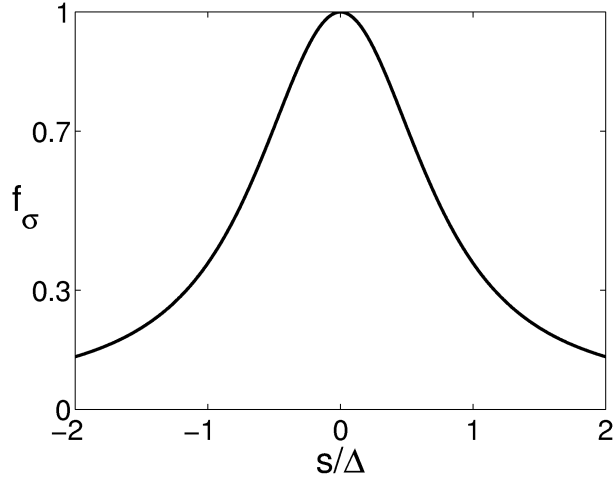}
\caption{Nonlinearity profile induced by the spatial overlap of the fundamental X-wave.\label{figurefsigma}
}
\end{figure}
$f_\sigma$, plotted in figure \ref{figurefsigma}, and such that $f_\sigma(0)=1$,
can be evaluated either numerically or analytically. In the latter case it can be expressed as 
\cite{noteA3}
\begin{equation}
\begin{array}{l}
  f_\sigma(\mu)=\frac{5}{512\mu^5}
\{\frac{4\mu(\mu^2-1)(3+14\mu^2+3\mu^4)}{(1+\mu^2)^2}
+3i(1+\mu^2)^2[
\log(-\frac{i}{2\mu})-\log(\frac{i}{2\mu})-
\log(\frac{i(-i+\mu)^2}{2\mu})+\log(-\frac{i(i+\mu)^2}{2\mu})]
\}\text{.}
\end{array}
\end{equation}

Therefore, $\sigma(s)$ is a bell-shaped function. Observing that its spatial extension is of the order
of $\Delta$, when the on-axis temporal dynamics is dominated by the $c$ envelope
(i.e. the temporal width of $c$ is smaller that $\Delta$), it can be approximated by $\sigma(0)$: 
\begin{equation}
\label{sigma0simplified}
\sigma(0)=\frac{8\,k\,k''}{5\Delta^2}\cong\frac{1}{\mathcal{W}^2_0}\text{,}
\end{equation}
The last relation stems from the fact that the spatial spectrum is $\Delta k_\perp\cong1/\mathcal{W}_0\cong\sqrt{k\,k''}/\Delta$.
The effective NLS can thus be approximately rewritten as
\begin{equation}
\label{1DNLSsimplified}
i\frac{\partial\,c}{\partial\,z}+\frac{k''}{2}\frac{\partial^2\,c}{\partial\,s^2}
+\frac{k\,n_2}{n_0\mathcal{W}_0^2}|c|^2\,c=0\text{,}
\end{equation}
so that the effective nonlinear coefficient
 has an additional factor, roughly given by $1/\mathcal{W}_0^2$ (or equivalently 
$\Delta k_\perp^2$), if compared to the plane wave value.
This shows the compensation of diffraction owing to the X-waves,
which behave as ``modes of free space''. $\sigma(s)$ resembles the
mode-overlap in a waveguide and the corresponding support to the 
nonlinear response along propagation. This holds valid as long as $z<<z_X$, i.e. as long as 
the nonlinearity is not averaged out by the ``sliding'' of the X-waves.

The fundamental soliton (when $\sigma(s)\cong\sigma(0)$) can be expressed in terms
of the peak $c_0^2$ of the energy distribution function; it reads
\begin{equation}
\label{integratesech}
A=\int c_0\, \text{sech}(\sqrt{\frac{c_0^2 \,k\,n_2\sigma(0)}{k''\,n_0}}s)\
exp[i\frac{c_0^2\,k\,n_2\sigma(0)}{2n_0}z]\xi_\slash^{(0)}(r,t,z,s)ds\text{,}
\end{equation}
which, using (\ref{AVscSemplified}) and (\ref{sigma0simplified}), can be approximated by 
\begin{equation}
\label{sechX}
A=2\pi\sqrt{k''}c_0 \,\text{sech}(\sqrt{\frac{c_0^2 \,k\,n_2}{\mathcal{W}_0^2 \,k''\,n_0}}t)\
exp(i\frac{c_0^2\,k\,n_2}{2n_0\mathcal{W}_0^2}z)\varphi^{(0)}(r,t)\text{.}
\end{equation}
As anticipated, from the expression of $\varphi_\slash^{(0)}$, 
the peak intensity $\mathcal{I}_0$ relates to $c_0$
by $c_0^2=\mathcal{I}_0\mathcal{W}_0^2$ (omitting inessential factors).

Eqs. (\ref{integratesech}) and (\ref{sechX}) show the ``dressing'' mechanism associated with the nonlinearity: the latter acts only on the shape of the envelope which, even in the
linear limit, would travel almost undistorted (as far as $z<<z_X$). 
Since this analysis can be repeated for any X-wave, nonlinear X-waves are not
numerable families of solutions (like multidimensional solitary waves), but have the same ``cardinality'' of the ``space'' of linear X-waves.

Conversely, in spite of this remarkable difference with solitary waves, the (approximate)
validity of an integrable model such as (\ref{1DNLSsimplified}) seems to establish a strong link 
with solitons. Clearly this analysis is far from being a rigorous settlement of this conclusion,
but could stimulate further research.
In the following section I will discuss some additional consequences of (\ref{1DNLS}) and (\ref{1DNLSsimplified}), 
pointing out non-trivial nonlinear dynamics of X-waves.
\section{Nonlinearly chirped X-waves and compression }
The fundamental X-wave/fundamental soliton discussed above is of limited
physical interest, because its invariant propagation is already ``embedded'' in its
linear counterpart.
If $C(\beta)$ is a very narrow function, or equivalently the $s-$width $\mathcal{T}_0$ of $c(s,z)$ is much wider than $\Delta$,
the on-axis profile of the 3D-beam is mainly determined by the X-wave.
This has been indicates as  \textit{the chirp regime}.

In Eq. (\ref{1DNLS}) the second derivative can be neglected, and the solution is essentially self-phase modulation:
\begin{equation}
\label{XSPM}
c(s,z)=A_0\exp[i\frac{k\,n_2}{n_0}\sigma(s)A_0^2 z]\text{.}
\end{equation}
Proceeding as before (\ref{XSPM}) yields the approximate \textit{nonlinearly chirped 
basis X-wave}
\begin{equation}
A\cong2\pi\sqrt{k}\sqrt{\frac{\mathcal{I}_0}{\mathcal{W}_0}} \varphi^{(p)}(r,t)
\exp[i\frac{k\,n_2\,z}{n_0\,\mathcal{I}_0}\,f_\sigma(t)].
\end{equation}
There is a fundamental difference between this type of chirp and what is typically
considered in fiber propagation, \cite{Agrawal} determined by the temporal power profile $|c|^2$. 
Even an idealized X-wave (corresponding to $C(\beta)$ proportional to a Dirac delta with respect to $\beta$,
and hence to a constant $c$) is chirped in a nonlinear medium because of its spatial profile (
reflected in $f_\sigma(s)$).

\begin{figure}
\includegraphics[width=80mm]{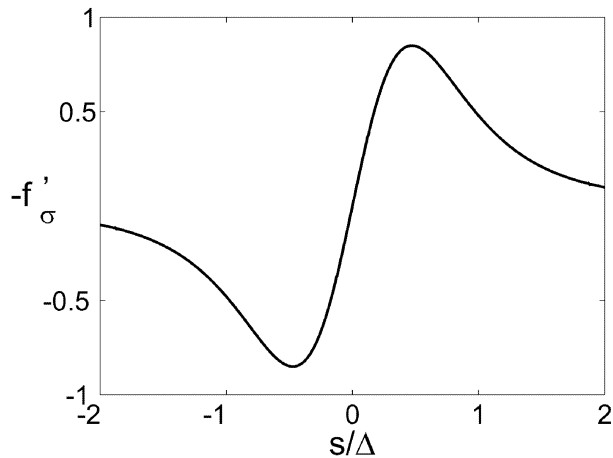}
\caption{Instantaneous frequency corresponding to X-wave self-overlap in Fig. \ref{figurefsigma}.\label{figurefreqshift}} 
\end{figure}

Since the earliest work on pulse splitting in normally dispersive Kerr media, 
it has been known that, at the initial stages, the on-axis 
temporal spectrum exhibits a double peak. \cite{Rothenberg92} Figure \ref{figurefreqshift} shows the instantaneous frequency 
\begin{equation}
\delta\omega=-\frac{k\,n_2\,z}{n_0\mathcal{I}_0}\frac{df_\sigma}{dt}\text{.}
\end{equation}
For higher order basis X-waves, more complicated spectral modulations can be expected. 
It is natural to identify the origin of this 
spectral splitting as a consequence of the self-phase modulation considered here.
It is also well stated that a chirped pulse may compress when propagating under 
 anomalous dispersion. With reference to the two frequencies at the
peaks in figure \ref{figurefreqshift},
they propagate according to Eq. (\ref{1DNLS}) with opposite velocities, such 
that the pulse gets compressed in the temporal domain. 
In the pulse splitting considered in \cite{Rothenberg92},
on-axis temporal compression immediately
follows the spectral splitting.

Thus, taking the whole process of pulse-splitting and admitting 
that an X-wave is formed in the initial stage (as in \cite{Conti03c}), the origin of the spectral splitting can be attributed to
self-phase modulation of an X-wave.
Nonlinear processes are therefore well suited for generating chirped X-waves, such as those 
recently addressed in \cite{Zamboni04}.

\section{Splitting and replenishment in Kerr media as a higher order soliton}
While the fundamental soliton and the chirp provide 
the simplest corollaries of Eq. (\ref{1DNLS}), the self-trapped behavior of the
3D beam is fundamentally due to the X-shape.
More interesting dynamics can be described referring
to multi-soliton solutions (obviously in \textit{the soliton regime}).   
The $N>1$ solitons \cite{Agrawal} are natural concepts in explaining 
splitting and replenishment, investigated numerically in 
\cite{Kolesik04} and experimentally in \cite{DiTrapani03b}.
Noteworthy, breathing linear X-waves have been reported. 
\cite{Shaarawi03, Zamboni03}

Even if the replenishment dynamics has been examined with reference to water
while including additional terms to Eq. (\ref{NLSdimensional3}), 
the most relevant features are somehow taken into account by this ``simple'' model, with
plasma formation and higher order dispersion playing a perturbative role. \cite{Kolesik04} 
Highly nonlinear phenomena, such as the shocks considered in \cite{Berge02}, 
are expected at very high fluences and will be neglected hereby.

For example, the N=2 soliton \cite{Agrawal} provides the approximate 
\textit{breathing nonlinear X-wave}
\begin{equation}
\label{breathingX}
A=2\pi\frac{k''}{\mathcal{T}_0}\sqrt{\frac{n_0}{k\,n_2\sigma(0)}}
u(\frac{z}{L_D},\frac{t}{\mathcal{T}_0})\varphi_\slash^{(0)}(r,t)\cong
-\frac{2k''\mathcal{W}_0}{\mathcal{T}_0}\sqrt{\frac{n_0}{n_2}}
u(\frac{z}{L_D},\frac{t}{\mathcal{T}_0})\frac{\Delta(t+i\Delta)}
{[(t+i\Delta)^2-k\,k''\,r^2]^{3/2}}\text{,}
\end{equation}
with
\begin{equation}
u(\xi,\tau)=4\frac{\cosh(3\tau)+3\exp(4i\xi)\cosh(\tau)}
{\cosh(4\tau)+4\cosh(2\tau)+3\cos(4\xi)}e^{i\xi/2}\text{.}
\end{equation}

\begin{figure}
\includegraphics[height=80mm]{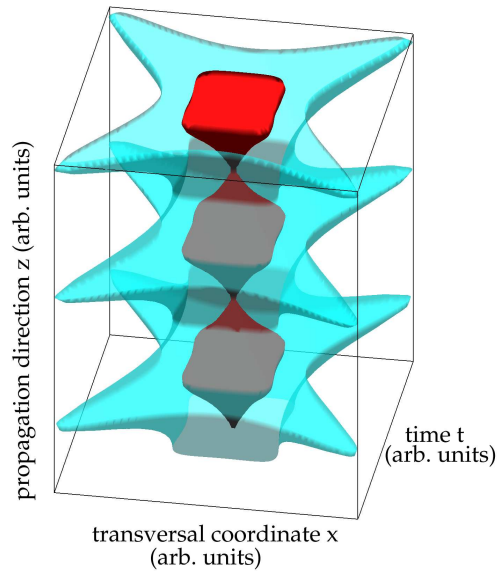}
\caption{(Color online) Typical spatio-temporal profile (at $y=0$) of a breathing X-wave. Two isosurfaces are displayed: the
darkest (red) corresponds to higher intensity.\label{FBX}}
\end{figure}

Figure \ref{FBX} shows a typical spatio-temporal profile obtained after Eq. (\ref{breathingX}). The periodic depletion and replenishment of the X-shaped distribution is apparent.
In propagation, the two-solitons, or breather-, solution pulsates and 
the beam evolves retaining most of its energy localized, but exhibiting the
non-trivial nonlinear dynamics of the X-wave.
The generation of a breather can also be thought at the origin
of splitting in the first stages of propagation. 
It is also noticeable that higher order solitons exhibit the spectral splitting
typically described in numerical simulations.

Before concluding, it is fruitful to summarize the picture of the splitting/replenishment phenomenon.
At the beginning a wide bell-shaped pulsed beam
 evolves into an X-wave, owing to the spatio-temporal pattern formation of X-wave instability \cite{Conti03c}. Then, an effective anomalous dispersion is experienced by 
the envelope of the finite-energy X-wave packet, and the \textit{chirp regime} is entered with
spectral splitting and on-axis compression.
Once the envelope width is sufficiently reduced, the \textit{soliton regime} starts, and the increased 
intensity through compression feeds the generation of a higher order soliton, or breather.
After some spatio-temporal oscillations, several mechanisms may intervene to stop the periodic behavior, e.g. 
losses (eventually of nonlinear origin, such as two-photon absorption) or simply that, for large propagation 
distances, the nonlinear response average out due to the
sliding between components of the finite energy X-wave packet.

\section{Conclusions}
Progressive undistorted waves, a generalization of the non-monochromatic
realm of self-invariant beams, seem to emerge as a valuable tool in 
numerous fields of applied research, from telecommunications to biophysics 
(supported by the recent experiments in water).
Their natural appearance in nonlinear processes, analyzed hereby 
in the optical domain but also available in acoustics as well as
in Bose Einstein condensation, can be considered a fundamental result.
The formation of X-waves during frequency generation appears as natural as the use of their paradigm in 
interpreting 3D+1 nonlinear dynamics.

While X-waves are known to carry infinite energy, and their superposition
has been previously employed to build finite energy solutions,
in this paper for the first time I have attempted to establish a general -albeit preliminary- theoretical framework for the extension  
of this approach to the nonlinear case.

The various steps of a basic nonlinear process, such as pulse-splitting
in normally dispersive media, can be revisited in terms of 
X-wave nonlinear dynamics: from the initial pulse compression 
to splitting/replenishment.
The results open the way to the investigation of elastic collisions between nonlinear X-waves, as
well as to establishing a link with parametric solitary waves in quadratic media.

According to standard textbooks, a soliton (or a bullet) is a non-perturbative solution
of a nonlinear wave equation. In this sense, the non dispersive and non diffracting wave-packets analyzed here cannot be 
considered solitons, because they exists even when a nonlinear susceptibility lacks.
Nevertheless, an intriguing connection between progressive undistorted waves and 
solitons seems to be at the inner stem of numerically and experimentally
investigated phenomena.
In this respect nonlinear X-waves can be considered  
a sort of \textit{chimera} of
modern nonlinear physics.
\section{acknowledgments}
I am indebted with 
 prof. G. Assanto (NOOEL) for the fruitful discussions and for a critical revision 
and with  
prof. S. Trillo (University of Ferrara) for his interest in this work.

\begin{thebibliography}{65}
\bibitem[{not({\natexlab{a}})}]{noteA1}
\bibinfo{note}{If $V$ is the velocity of the X-wave, and $V_g=1/k'$ is the
  group velocity at $\omega_0$ it is: $V=1/(\beta+k')$}.
\bibitem[{not({\natexlab{b}})}]{noteX}
\bibinfo{note}{In this manuscript the term \textit{X-wave} is defined by the
  mathematical expression. Different definitions can be found in the literature
  (see the cited references), however they are not directly useful in the
  presence of material dispersion, or of nonlinearity, or for \textit{envelope
  X-waves}}.
\bibitem[{not({\natexlab{c}})}]{noteA2}
\bibinfo{note}{In principle, even the harmonic generation by a pump X-wave,
  travelling at a given velocity could be considered.}
\bibitem[{not({\natexlab{d}})}]{noteA3}
\bibinfo{note}{Care is needed because of the use of the inverse functions and
  corresponding branch cuts. The use of the complex $\log(z)$ enables treating
  together the branches $\mu<0$ and $\mu>0$.}
\end{thebibliography}

%
\end{document}